\documentclass[prd,twocolumn,preprintnumbers,superscriptaddress,nofootinbib]{revtex4}
\usepackage[a4paper, hdivide={1.91cm,,1.165cm}, vdivide={1.83cm,,3.6cm}]{geometry}

\usepackage{amstext,amssymb}
\usepackage{amsmath}
\usepackage{graphicx}
\usepackage[hyperfootnotes=false]{hyperref}
\usepackage{xspace}
\usepackage{color}
\usepackage{slashed} 
\usepackage{multirow}
\usepackage[normalem]{ulem}

\begin{document}
\title{Unifying the flavor origin of dark matter with leptonic nonzero $\theta_{13}$}
\author{Subhaditya Bhattacharya}
\email{subhab@iitg.ernet.in}
\affiliation{Department of Physics, Indian Institute of Technology Guwahati, Assam-781039, India}
\author{Biswajit Karmakar}
\email{k.biswajit@iitg.ernet.in}
\affiliation{Department of Physics, Indian Institute of Technology Guwahati, Assam-781039, India}
\author{Narendra Sahu}
\email{nsahu@iith.ac.in}
\affiliation{Department of Physics, Indian Institute of Technology Hyderabad, Yeddumailaram, 502205, 
Telengana, India}
\author{Arunansu Sil}
\email{asil@iitg.ernet.in}
\affiliation{Department of Physics, Indian Institute of Technology Guwahati, Assam-781039, India}
\begin{abstract}
We propose a flavor symmetric approach to unify the origin of dark matter (DM) with the non-zero $\theta_{13}$ in
the lepton sector. In this framework, the breaking of a $U(1)$ flavor symmetry to a remnant $Z_2$ ensures
the stability of the DM and gives rise to a modification to the existing $A_4$-based tri-bimaximal neutrino mixing
to attain the required non-zero values of $\sin \theta_{13}$. This results in a range of Higgs portal coupling of 
the DM which can be potentially accessible at various ongoing and future direct and collider search experiments.
\end{abstract}

\maketitle
Flavor symmetries play important roles in understanding many issues in particle physics including quark and lepton 
mixing as well as mass hierarchies. Historically a global $U(1)$ flavor symmetry was proposed to explain the quark 
mass hierarchy and Cabibbo mixing angle~\cite{Froggatt:1978nt} which was extended to explain neutrino masses and 
mixing later. Among others, a tri-bimaximal (TBM) lepton mixing generated from a discrete flavor symmetry such as
$A_4$ gets particular attention~\cite{Ma:2001dn,Altarelli:2005yx} due to its simplicity and predictive nature. 
However, the TBM mixing primarily is associated with a vanishing reactor mixing angle $\theta_{13}$ which is against
the recent robust observation of non-zero $\theta_{13} \approx 9^{\circ}$~\cite{Capozzi:2013csa,Gonzalez-Garcia:2014bfa, Forero:2014bxa}
by DOUBLE CHOOZ~\cite{Abe:2011fz}, Daya Bay~\cite{An:2012eh}, RENO~\cite{Ahn:2012nd} and T2K~\cite{Abe:2013hdq} 
experiments. Hence, an alteration to TBM structure has been under scanner.  

Understanding the nature of dark matter (DM) is another outstanding problem in particle physics today. Although
astrophysical evidences, such as rotation curves of galaxies, gravitational lensing and large scale structure of
the Universe supports the existence of DM ~\cite{DM_review}, a discovery in laboratory is still awaited. The 
relic abundance of DM has been measured by WMAP~\cite{wmap} and PLANCK~\cite{planck} satellite experiments to
be about 26.8\% of the total energy budget of the Universe. Although this hints towards a broad classification 
of DM scenarios, its properties apart from gravitational interactions, are not known yet. 

\begin{figure}[ht]
 $$
\includegraphics[height=4.5cm]{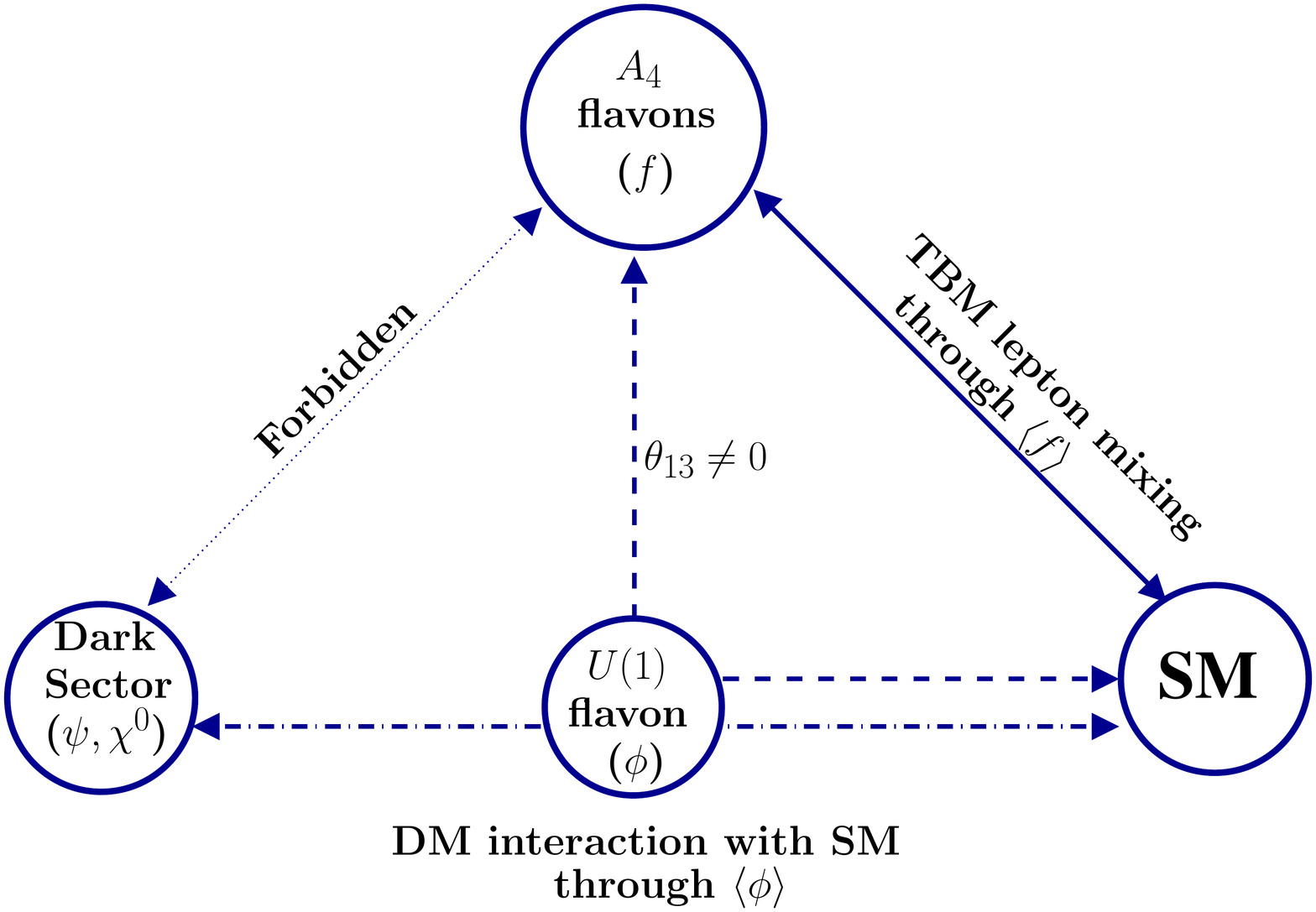}
$$
  \caption{Non-zero values of $\sin \theta_{13}$ predict Higgs portal couplings of DM via a $U(1)$ flavour symmetry: a schematic presentation.}\label{fig:schm}
\end{figure}

In this paper we propose a $U(1)$ flavor extension of the Standard Model (SM) to unify the origin of DM with 
the simultaneous realization of non-zero $\sin \theta_{13}$ in the lepton sector. For this purpose, we presume
the existence of a TBM neutrino mixing pattern (in a basis where charged leptons are diagonal) and a dark sector 
consisting of vector like leptons. We will argue that this serves as a minimal extension of the SM to accommodate DM and
non-zero $\sin \theta_{13}$. A pictorial presentation of the model is shown in Fig.\ref{fig:schm}. Here $f$
represents the flavon fields charged under $A_{4}$, the vacuum expectation values (vevs) of which 
($\langle f \rangle$) would break the $A_4$ and generate the flavor structure of the lepton sector. 
The flavon field ($\phi$) charged under the $U(1)$ plays the role of a messenger between 
the dark sector and SM particles including left-handed neutrinos. The $U(1)$ symmetry, once allowed to be broken by the 
vev of $\phi$, generates a non-zero $\sin \theta_{13}$ and a Higgs portal coupling to the vecor-like leptonic DM in 
the effective field theory. We will show that the non-zero values of $\sin \theta_{13}$ are correlated to 
the Higgs portal coupling of the DM which yields the correct relic density measured by WMAP~\cite{wmap} and PLANCK~\cite{planck}. 
Future direct search experiments, such as Xenon1T~\cite{Aprile:2015uzo} and the Large Hadron Collider (LHC)~\cite{Bhattacharya:2015qpa,Arina:2012aj} 
can establish a bridge between the two invisible sectors by measuring the Higgs portal coupling of DM. 


%

We consider an effective field theory approach for the demonstration purpose and begin by assuming a typical
well known structure of the neutrino mass matrix~\cite{Ma:2001dn,Altarelli:2005yx}, $\left( m_{\nu}\right)_0$,
given by
\begin{eqnarray}\label{neutrino-mass}
 (m_{\nu})_0=\left(
\begin{array}{ccc}
      a-2b/3   &b/3    &b/3\\
       b/3    &-2b/3    &a+b/3\\
       b/3    &a+b/3   &-2b/3
\end{array}
\right),
\end{eqnarray}
which results in a TBM neutrino mixing pattern while the charged lepton mass matrix is diagonal. The TBM mixing 
matrix~\cite{Harrison:1999cf} can be represented by:
 \begin{eqnarray}\label{utb}
 U_{TBM}=\left(
\begin{array}{ccc}
 \sqrt{\frac{2}{3}} & \frac{1}{\sqrt{3}} & 0 \\
 -\frac{1}{\sqrt{6}} & \frac{1}{\sqrt{3}} & -\frac{1}{\sqrt{2}} \\
 -\frac{1}{\sqrt{6}} & \frac{1}{\sqrt{3}} & \frac{1}{\sqrt{2}}
\end{array}
\right), 
\end{eqnarray}
implying $\sin \theta_{13}=0$, $\sin^2 \theta_{12}=1/3$ and $\sin^2 \theta_{23}=1/2$. The above structure
of $(m_{\nu})_0$ can be obtained in a $A_4$ based set-up either in a type-I or II see-saw 
framework~\cite{Karmakar:2014dva,Branco:2012vs,Karmakar:2015jza} or through higher dimensional lepton number violating
operators. For example, we can have Altarelli-Feruglio (AF) model~\cite{Altarelli:2005yx}, where the SM doublet leptons
($\ell$) are transforming as triplet under the $A_4$ while the singlet charged leptons $e_R, \mu_R$ and $\tau_R$ transform
as $1, 1^{''}$ and $1^{'}$ respectively. Then a higher dimensional operator of the form, 
$(\ell H \ell H)(\xi- y \phi_S)/\Lambda^2$ can be considered, where $\xi$ and $\phi_S$ are singlet and triplet flavon 
fields (they are SM singlet and transform under $A_4$) respectively. $\Lambda$ is the cut off scale of the theory and
$y$ represents the relative strength between the two couplings involved. Once these flavons get vevs, a flavor structure
can be generated after electroweak symmetry breaking with $a=(v^2/\Lambda)\epsilon$ and $b=y(v^2/\Lambda)\epsilon$ where
$\epsilon=\langle \xi \rangle / \Lambda=\langle \phi_S \rangle / \Lambda$. With a judicious choice 
of additional discrete symmetries like $Z_3$ or more, one can ensure that no other terms involving these flavons and SM
fields are allowed at $1/\Lambda^2$ order or below so as to keep the structure of $(m_{\nu})_0$ intact as in 
Eq. (\ref{neutrino-mass}). In what follows, we introduce an additional global $U(1)$ flavor symmetry which will be 
broken into a remnant $Z_2$ and an additional contribution to $(m_{\nu})_0$ becomes functional. None of the fields in the above dimension-6 operators, responsible for TBM mixing, would carry any $U(1)$ charge in order to generate non-zero $\sin\theta_{13}$ and establish a connection to the dark sector.

Simplest way to connect non-zero $\sin\theta_{13}$ and dark sector is achieved by assuming a minimal fermionic DM framework consisted of a vector-like $SU(2)_L$ doublet fermion $\psi^T=(\psi^0,\psi^-)$ and a vector-like neutral singlet fermion $\chi^0$~\cite{Bhattacharya:2015qpa}. These fermions are charged under the additional $U(1)$ flavor symmetry, but neutral under the existing symmetry in the neutrino sector 
(say the non-abelian $A_4$ and additional discrete symmetries required). We also introduce two other SM singlet flavon fields $\phi$ and $\eta$ which carry equal and opposite charges under the $U(1)$ symmetry but transform as $1$ and $1^{'}$ under 
$A_4$. Note that the SM fields are neutral under this additional $U(1)$ symmetry. The effective Lagrangian, invariant under the symmetries considered, describing the interaction between the dark and the SM sector is then given by: 
\begin{equation}\label{lagrangian}
 \mathcal{L}_{\rm int}=\left(\frac{\phi}{\Lambda}\right)^n 
\overline{\psi}\widetilde{H}\chi^0
  +\frac{(\ell H \ell H)\phi\eta}{\Lambda^3}\,.
\end{equation}
We keep $n$ as a free parameter at present. The first term is allowed since the $U(1)$ charge of $\phi^n$ is compensated by
$\psi$ and $\chi^0$, while the second term is allowed since the $U(1)$ charges of $\phi$ and $\eta$ cancel with each other.
This also ensures that $\phi$ and $\eta$ do not take part in $(m_{\nu})_0$. The detailed structure of the scenario is left 
for a future work~\cite{all_authors}. The idea of introducing a vector like fermion in the dark sector is also motivated by
the fact that we expect a replication of the SM Yukawa type interaction to be present in the dark sector as well. Here the
$\phi$ field plays the role of the messenger field similar to the one considered in~\cite{Calibbi:2015sfa}. 
See~\cite{darkA4prev}, for some earlier efforts to relate $A_4$ flavor symmetry to DM.

When $\phi$ and $\eta$ acquire vevs, the $U(1)$ symmetry breaks into a remnant $Z_2$ symmetry under which the vector-like fermions $\psi$ and $\chi^0$ are odd. Consequently the DM emerges as an admixture of the neutral component of the vector-like fermions $\psi$ and $\chi^0$ and yields a larger region of allowed parameter space as we will shortly demonstrate. The interaction strength of the DM with the SM Higgs is then given by $\left( \langle \phi \rangle/\Lambda \right)^n \equiv \epsilon^n$. Similarly the second term in Eq. (\ref{lagrangian}) provides an additional contribution to the light neutrino mass matrix as follows: 
\begin{equation}\label{neutrino-correction}
\delta m_\nu= \left(
\begin{array}{ccc}
      0   &0   &d\\
      0    &d   &0\\
      d    &0  &0
\end{array}
\right), 
\end{equation} 
where $d=(v^2/\Lambda)\epsilon^2$ with $\epsilon=\langle \phi \rangle/ \Lambda \equiv \langle \eta \rangle /\Lambda$.
This typical flavor structure follows from the involvement of $\eta$ field, which transforms as $1'$ under $A_4$~\cite{Shimizu:2011xg}.

From Eqs. (\ref{neutrino-mass}) and (\ref{neutrino-correction}), we get the light neutrino mass matrix as $m_\nu=(m_\nu)_0 + \delta m_\nu$. We have already seen that the $(m_\nu)_0$ can be diagonalized by $U_{TBM}$ alone, so an additional rotation ($U_1$) is required to diagonalize $m_\nu$: 
\begin{eqnarray}\label{u1}
U_1 =\left(
\begin{array}{ccc}
 \cos\theta_{\nu}             & 0 & \sin\theta_{\nu}  \\
     0                    & 1 &            0 \\
 -\sin\theta_{\nu}  & 0 &        \cos\theta_{\nu} 
\end{array}
\right)\,. 
\end{eqnarray}
Here we assume all parameters $a, b, d$ are real for simplicity. We therefore obtain~\cite{Karmakar:2014dva}
\begin{equation}
 \tan 2 \theta_{\nu}  = \frac{\sqrt{3} d}{d - 2a} = \frac{\sqrt{3}\epsilon}{\epsilon-2}.
\end{equation}
Then comparing the standard $U_{PMNS}$ parametrization and neutrino mixing matrix $U_{\nu}(=U_{TBM}U_1)$ we get
$
\sin\theta_{13}=\sqrt{\frac{2}{3}}\left|\sin\theta_{\nu} \right|,
\sin^2\theta_{12}=\frac{1}{3(1-\sin^2\theta_{13})},
 \sin^2\theta_{23}=\frac{1}{2}
+\frac{1}{\sqrt{2}}\sin\theta_{13},
\delta={\rm arg}[(U_1)_{13}]=0
$. 

\begin{figure}[!h]
$$
\includegraphics[height=4.5cm]{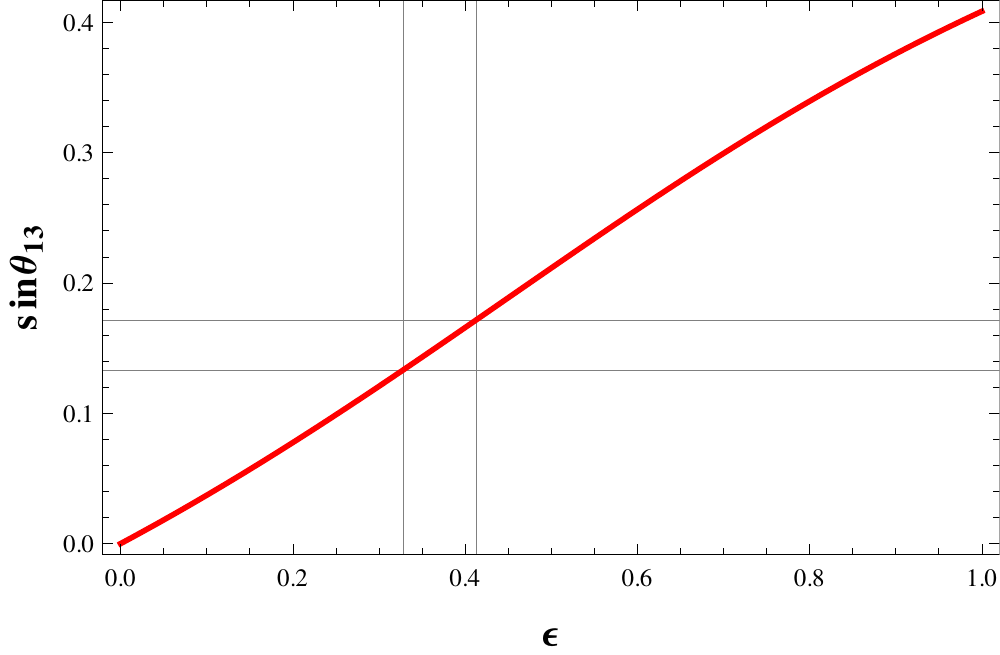}
$$
\caption{ Plot of $\sin\theta_{13}$ against $\epsilon$. Here 3$\sigma$ range~\cite{Forero:2014bxa} for 
$\sin\theta_{13}$ fixes $\epsilon$ in the range 0.328-0.4125. }
\label{fig:s}
\end{figure}

Clearly $\sin\theta_{13}$ depends only on $\epsilon$ as shown in Fig \ref{fig:s}. The horizontal patch in 
Fig \ref{fig:s} denotes the allowed 3$\sigma$ range of $\sin\theta_{13}$ ($\equiv$ 0.1330-0.1715)~\cite{Forero:2014bxa}
which is in turn restrict the range of $\epsilon$ parameter denoted by the vertical patch on the figure. 
Note that the interaction strength of DM with the SM particles depend on $\epsilon^n$. Therefore we 
find that the size of $\sin\theta_{13}$ is intimately related with the Higgs portal coupling of DM. 
This is an important observation in this letter and is demonstrated in the rest of the paper. The two other 
mixing angles $\theta_{12}$ and $\theta_{23}$ fall in the right ballpark while light neutrino mass satisfy the 
$\Delta m^2_{\odot}=7.60\times10^{-5}$ eV$^2$ and $|\Delta m^2_{atm}|=2.48\times10^{-3}$
eV$^2$~\cite{Capozzi:2013csa,Gonzalez-Garcia:2014bfa, Forero:2014bxa}. 

Now we focus on the 1st term of Eq. (\ref{lagrangian}) to estimate the relic density of dark matter as a function of $\epsilon$. Note, both flavour and the dark sector constrains the ratio $\epsilon$, instead of the new physics scale $\Lambda$ in the effective operator formalism considered here.
Since $\psi$ and $\chi^0$ are vector-like fermions, they can have  bare masses, $M_\psi \overline{\psi} \psi$ and $M_\chi \overline{\chi^0}\chi^0$, which are not protected by the SM
symmetry\footnote{There are two additional terms, $\overline{\chi^0} \chi^0 H^{\dagger} H/\Lambda$ and $\overline{\psi} \psi H^{\dagger}
H/{\Lambda}$, which are also allowed by the symmetry considered. However their contribution ($\sim v^2/\Lambda$)
to the mass matrix $\mathcal{M}$ is negligibly small compared to the bare masses and $Yv$. They also
have negligible impact on the DM annihilation processes as the DM-DM-$h$ vertex would be suppressed by
$v/\Lambda$.}.
The electroweak phase transition also gives rise a mixing between $\psi^0$ and $\chi^0$. In the basis $(\chi^0, \psi^0)$, the mass matrix is given by
\begin{equation}
\mathcal{M} = \begin{pmatrix}  M_\chi & Yv\cr \\
Yv & M_\psi \,,
\end{pmatrix}
\end{equation} 
where $Y=\epsilon^n$. Diagonalizing the above mass matrix we get the mass eigenvalues as $M_1$ and $M_2$ corresponding
to the mass eigenstates $\psi_1$ and $\psi_2$. We assume that $\psi_1$ is the lightest odd particle and hence constitute
the DM of the Universe. The mixing angle is given by 
\begin{equation}\label{theta-d}
\sin 2\theta_d \simeq \frac{2 Y v}{\Delta M}\,, 
\end{equation}
in the small mixing limit where $\Delta M= M_2-M_1$. Here, we note that small mixing is necessary for the model to provide
a DM with viable relic density. This is because, the larger is the doublet content in DM $\psi_1$, the annihilation goes up significantly in particular through $\psi_1 \overline{\psi_1} \to W^{+}W^{-}$. So in the limit, $\psi_2$ is dominantly a doublet having a small admixture of the singlet one. This implies that $\psi_2$
mass is required to be larger than 45 GeV in order not to conflict with the invisible $Z$-boson decay width. In the physical spectrum we also have a charged fermion $\psi^+ (\psi^-)$ with mass $M^+ (M^-) = M_1 \sin^2 \theta_d + M_2 \cos^2 \theta_d$. In the limit $ \theta_d \to 0$, $M^\pm =M_2= M_\psi$.

The relic density of the $\psi_1$ dark matter is mainly dictated by annihilations $\overline{\psi_1}\psi_1 \to W^+W^-$ through $SU(2)$ gauge coupling and
$\overline{\psi_1} \psi_1 \to h h$ through Yukawa coupling introduced in Eq. \ref{lagrangian}. The other relevant channels are mainly co-annihilation of $\psi_1$ with $\psi_2$ and $\psi^\pm$ which dominantly contribute to relic density in a large parameter space~\cite{griest,Bhattacharya:2015qpa,Cynolter:2008ea,
Cohen:2011ec,Cheung:2013dua}. The dark-sector is mainly dictated by three parameters $\sin\theta_d, M_{1}, M_{2}$. 
In the following we use Micromega~\cite{Belanger:2008sj} to find the allowed region of correct relic abundance for $\psi_1$ DM 
satisfying WMAP~\cite{wmap} constraint~\footnote{The range we use corresponds to the WMAP results; the PLANCK 
constraints $0.112 \leq \Omega_{\rm DM} h^2 \leq 0.128$~\cite{planck}, though more stringent, do not lead to significant changes
in the allowed regions of parameter space.}
\begin{equation}
0.094 \leq \Omega_{\rm DM}  h^2 \leq 0.130 \,.
\label{eq:wmap.region} 
\end{equation}

\begin{figure}[thb]
$$
\includegraphics[height=5.5cm]{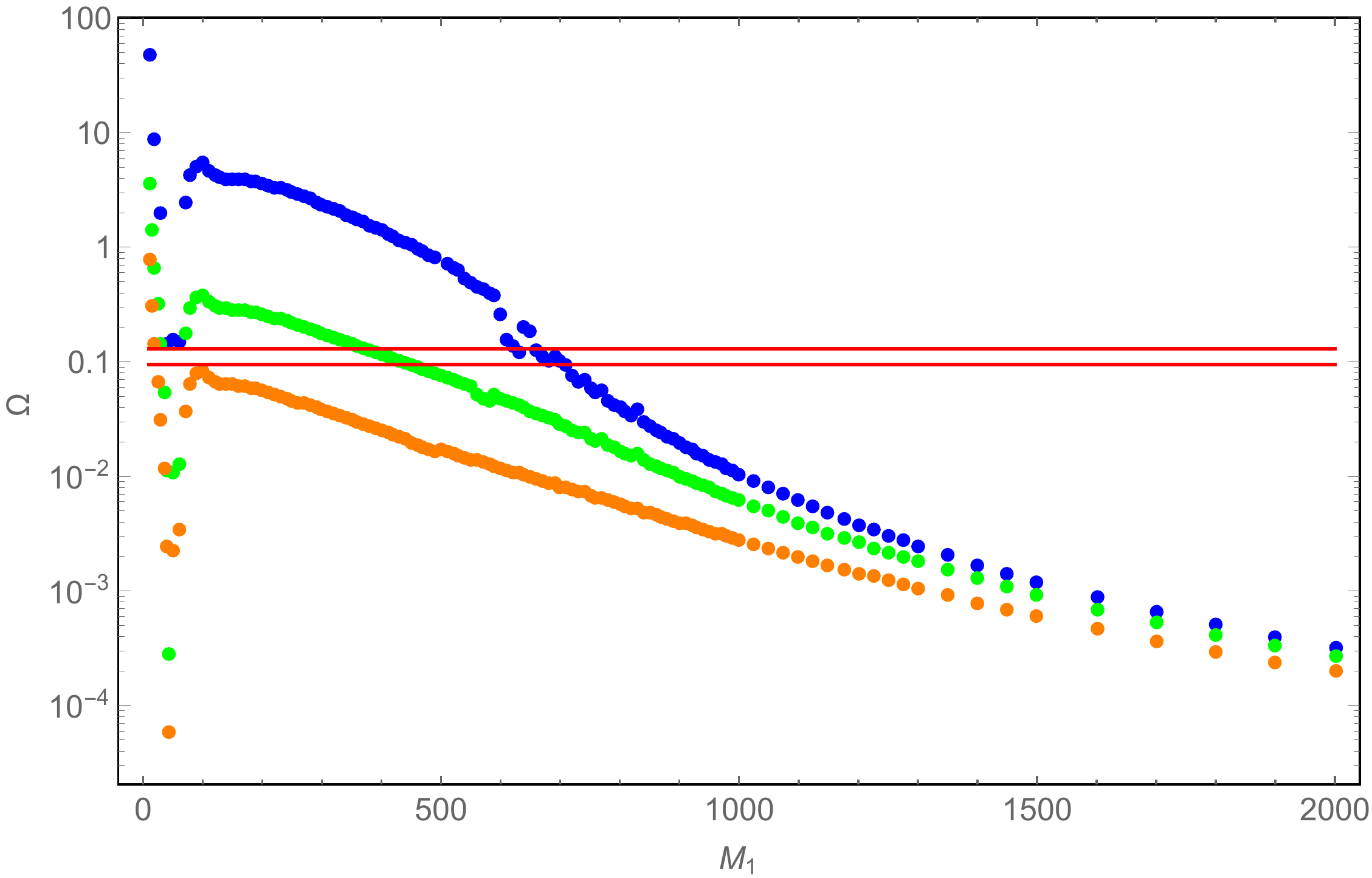}
$$
\caption{Relic density vs DM mass $M_1$ (in GeV) for different choices of $Y=0.04,0.08,0.115$ (Blue, Green, Orange respectively)
with $\Delta M = 100$ GeV. Horizontal lines define the correct relic density.}
\label{fig:relic1}
\end{figure}

\begin{figure}[thb]
$$
\includegraphics[height=5.5cm]{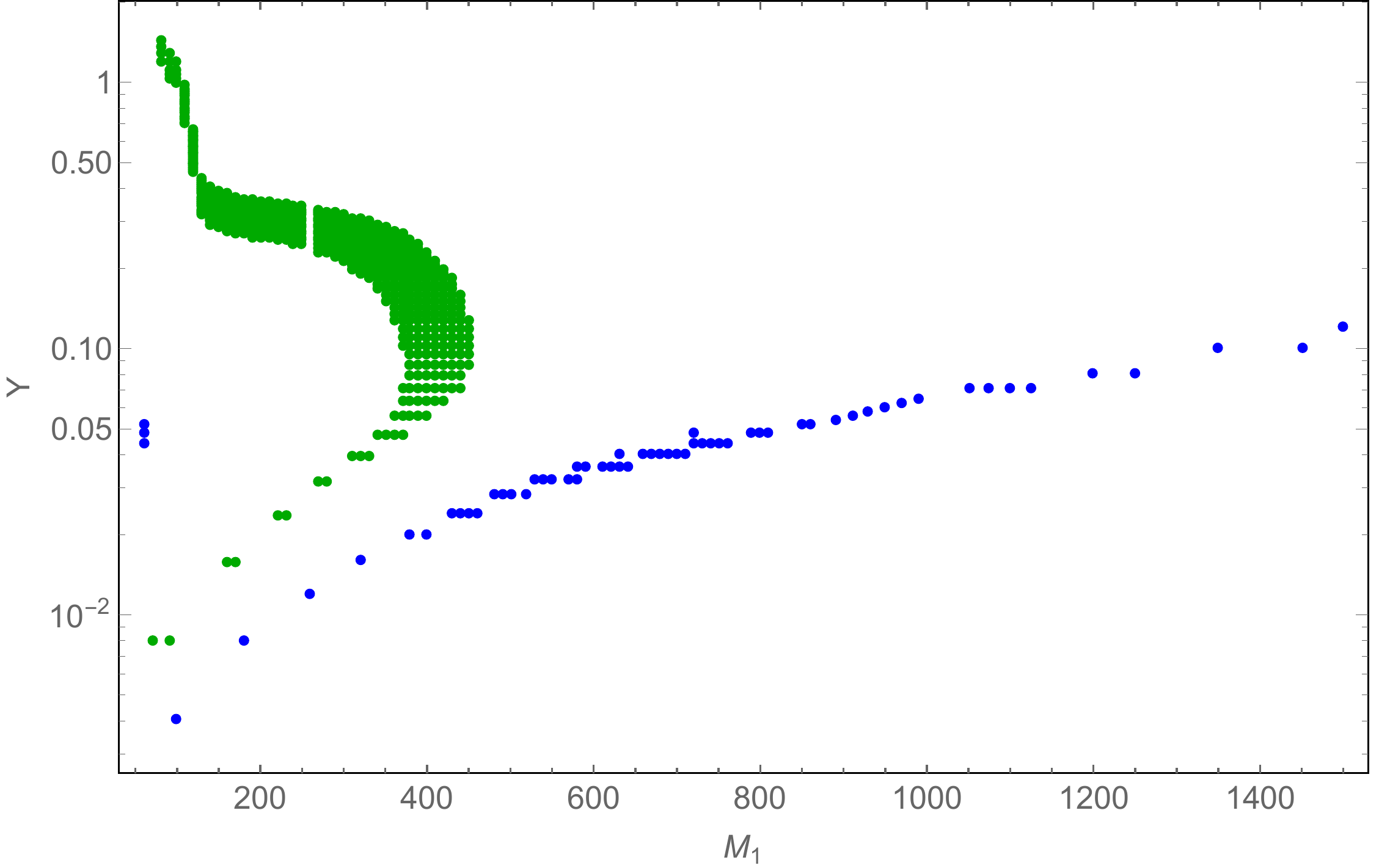}
$$
\caption{$Y$ versus $M_1$ (in GeV) for correct relic density (Eq. \ref{eq:wmap.region}). 
$\sin \theta=0.1,~0.2$ (Blue and Green respectively) has been chosen, while $\Delta M$ vary arbitrarily.} 
\label{fig:Y-M1}
\end{figure}

In Fig. \ref{fig:relic1} we plot relic density versus DM mass $M_1$ for different choices of $Y$ (= 0.04, 0.08, 0.115 with Blue,
Green and Orange respectively from top to bottom), keeping the mass difference $\Delta M$ fixed at 100 GeV. The mixing angle
$\sin\theta_d$ (as obtained using Eq. (\ref{theta-d})) associated with the top (Blue) line in 0.1 and it increases to 0.2 and 
0.3 for the middle (Green) and bottom (Orange) lines. As the mixing increases, the doublet component starts to dominate and
hence give larger cross-section which leads to a smaller DM abundance. Note that $\sin\theta_d=0.3$ ($Y=0.115$ with $\Delta M= 100$ GeV) can barely 
satisfy relic density, where annihilations through $Z$ mediation becomes large. In Fig. \ref{fig:Y-M1}, we plot $Y$ versus
$M_1$ to produce correct relic density with $\sin\theta_d=0.1$ and 0.2 while varying $\Delta M$. It points out a
relatively wide DM mass range satisfy the relic density constraint. It also shows that for $\sin\theta_d=0.1$ (generally true 
for $\sin\theta_d \le 0.1$), the annihilations are never enough to produce correct density and co-annihilations play a crucial 
part resulting the blue curve rising with the DM mass. For $\sin\theta_d=0.2$ (green patch), smaller DM mass regions get 
contributions from coannihilation with small $Y$ and annihilations only for large $Y$, while the region close to DM mass 
400 GeV has a significant contribution from $Z$ mediation. 

\begin{figure}[thb]
$$
\includegraphics[height=5.5cm]{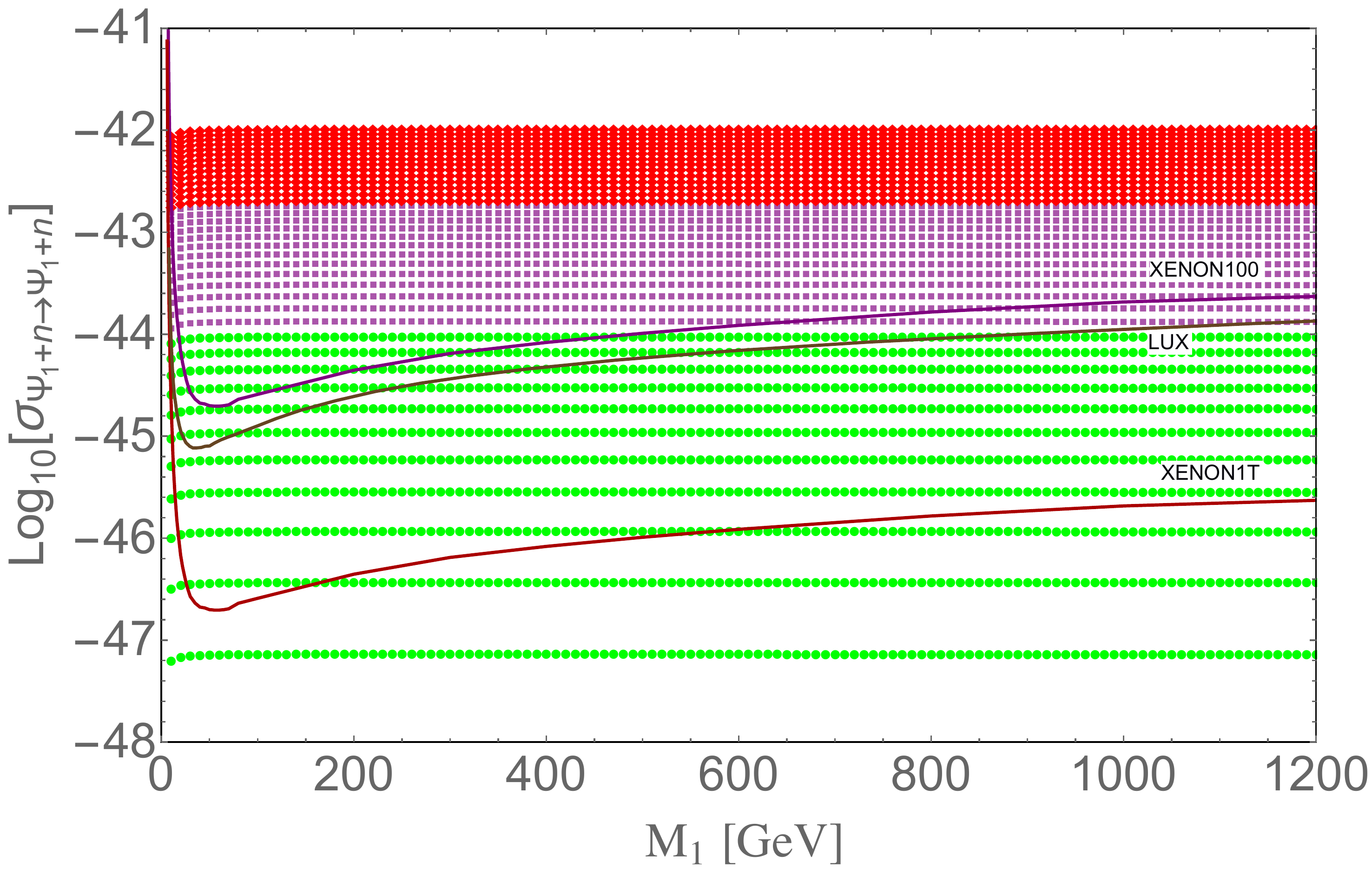}
$$
\caption{Allowed values of the Higgs portal coupling $Y$ by the direct search experiments, Xenon100,
LUX and Xenon-1T: $Y:\{0.001-0.05\}$ (Green), 
$Y:\{0.05-0.1\}$ (Purple), $Y:\{0.1-0.15\}$ (Red). $\Delta M= 100$ GeV is used for the scan.}
\label{fig:DD}
\end{figure}

The most stringent constraint on the Higgs portal coupling $Y \simeq \sin 2\theta_d \Delta M/(2v)$ comes from the
direct search of DM at Xenon-100~\cite{Aprile:2012nq}, LUX~\cite{Akerib:2013tjd} as demonstrated in Fig.\ref{fig:DD}. 
We see that the bound from LUX, constraints the coupling: $Y \sim 0.05$ for DM masses $\gtrsim 800$ GeV (Green points).
The Yukawa coupling needs to be even smaller for $M_1 \simeq 100$ GeV. Though large couplings are allowed by correct
relic density, but they are highly disfavored by the direct DM search at terrestrial experiments.  Note that these 
constraints are less dependent on $\Delta M$ as to the mixing angle, which plays otherwise a crucial role in the 
relic abundance of DM.

\begin{figure}[thb]
$$
\includegraphics[height=4.7cm]{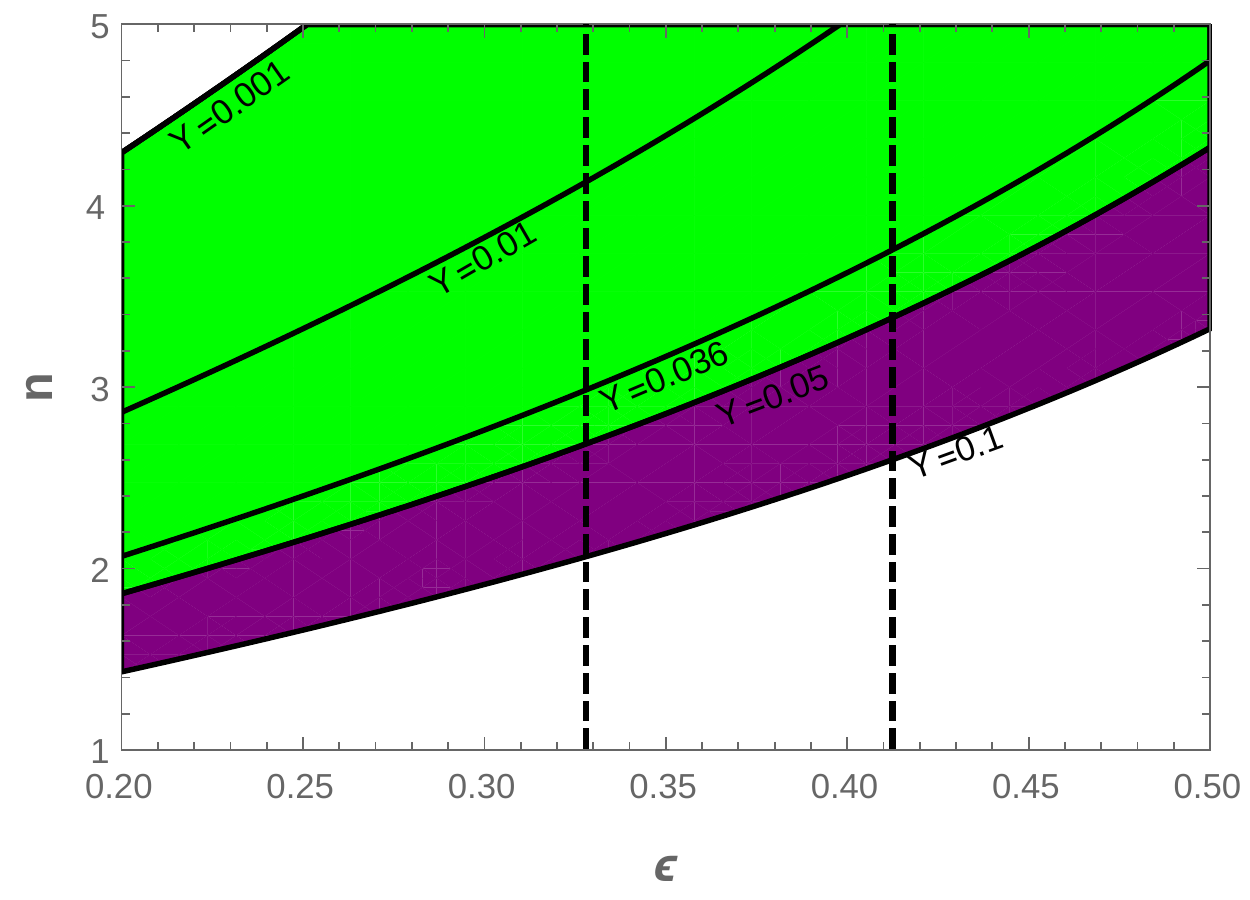}
$$
\caption{ $n$ vs $\epsilon$ to generate different values of $Y=\epsilon^n$.}
\label{fig:epsilon}
\end{figure}

 We can now combine the outcome of the two sectors into Fig. \ref{fig:epsilon}. The allowed range of $Y$-values
 can be translated in terms of $\epsilon-n$ as shown here. Correct $\sin\theta_{13}$ allowed $\epsilon$ within 
 $0.328 - 0.4125$ (see Fig. (\ref{fig:s}). Therefore, the Higgs portal couplings: $Y\lesssim 0.05$, allowed by
 correct relic density and direct search of DM can be satisfied with $n=3$ or more. 

The $U(1)$ symmetry of the model is broken by the vev of a flavon field to a remnant $Z_2$, 
whereas the breaking of $A_4$ (and additional discrete symmetry) is responsible for producing 
the flavor structure of neutrino mass matrix. The details of symmetry breaking pattern 
and charge assignment of the flavon fields is worthy of attention. Non-zero $\sin \theta_{13}$ 
appeals for finite values of phases and hence CP-violations, which have been ignored in this letter. 
They will be discussed together in a future publication~\cite{all_authors}.   

In summary, the observed value of non-zero $\sin \theta_{13}$ and its link to Higgs portal coupling of a vector-like
fermionic DM was obtained in a further $U(1)$ flavor extension of the SM. We showed that the non-zero values of 
$\sin \theta_{13}$ fixes a range of Higgs portal coupling $Y = \epsilon^n, n\gtrsim3$  which can be probed at the
future direct DM search experiments such as Xenon-1T.  Also note that the next to lightest stable particle (NLSP)
could be a charged fermion which can be searched at the LHC~\cite{Arina:2012aj}. In the limit of small $\sin\theta_d$,
the NLSP can give rise to a displaced vertex at LHC, a rather unique signature for the model discussed~\cite{Bhattacharya:2015qpa}.

\section{Acknowledgements}
The work of SB is partially supported by DST INSPIRE grant no PHY/P/SUB/01 at IIT Guwahati. 
NS is partially supported by the Department of Science and Technology, Govt. of India under 
the financial Grant SR/FTP/PS-209/2011. 


\end{document}